\documentclass{article}
\usepackage{algorithm}
\usepackage{algorithmic}
\usepackage{graphicx}    % 处理图片
\usepackage{subcaption}  % 支持子图
\usepackage[english]{babel}
\usepackage[
  a4paper,
  top=2cm,
  bottom=2cm,
  left=3cm,
  right=3cm,
  marginparwidth=1.75cm
]{geometry}

\usepackage{indentfirst} % 强制首段缩进
\usepackage{ragged2e} % 两端对齐
\usepackage{amsfonts} % 数学符号

\justifying % 两端对齐
\usepackage{amsmath} % 数学公式
\usepackage{hyperref} % 超链接（放置在较后位置避免冲突）
\usepackage{natbib} % 参考文献

\title{CRMD: Complex Robust Modal Decomposition}

\author{
  Wang Hao$^{*}$ \quad
  \href{mailto:25b905039@stu.hit.edu.cn}{25b905039@stu.hit.edu.cn} \\
  Kuang Zhang \quad \href{mailto:zhangkuang@hit.edu.cn}{zhangkuang@hit.edu.cn} \\
  Hou Chengyu$^{\dagger}$ \quad \href{mailto:houcy@hit.edu.cn}{houcy@hit.edu.cn} \\
  Tan Chenxing \quad \href{mailto:3424611356@qq.com}{3424611356@qq.com} \\
   Cui Weiming\quad \href{mailto:hitwmy@126.com}{hitwmy@126.com} \\
  Fu Weifeng \quad \href{mailto:2298000456@qq.com}{2298000456@qq.com} \\
  Yao Xingran\quad \href{mailto:2220205413@qq.com}{2220205413@qq.com} \\
  \small School of Electronics and Information Engineering, \\
  \small Harbin Institute of Technology, Harbin 150001, China
}

\begin{document}
\maketitle % 生成标题页，避免手动换行导致的空白
\thanks{$^{*}$ First author. $^{\dagger}$ Corresponding author: houcy@hit.edu.cn.}

  \begin{abstract}
   Compared to real-valued signals, complex-valued signals provide a unique and intuitive representation of the phase of real physical systems and processes, which holds fundamental significance and is widely applied across many fields of science and engineering. In this paper, we propose a robust modal decomposition (RMD) in the complex domain as a natural and general extension of the original real-valued RMD. We revisit and derive the mathematical principles of RMD in the complex domain, and develop an algorithmic version tailored for this domain. Extensive experiments are conducted on synthetic simulation datasets and real-world datasets from diverse fields, including a millimeter-wave radar physiological signal detection dataset, a faulty bearing dataset, a radio-frequency unmanned aerial vehicle identification dataset, and a WiFi CSI-based respiration detection dataset. The results demonstrate that the proposed complex-domain robust modal decomposition significantly improves performance across these various applications.
  \end{abstract}

  \noindent

  \textbf{Keywords:} Complex Modal Decomposition, Complex Eigenvalue Decomposition, Constrained
  Bandwidth, Robust Modal Decomposition

  \section{INTRODUCTION}

  It is well known that the phase of a signal is a relative quantity, whereas conventional real-valued signals can only provide a description of the instantaneous amplitude, lacking a description of phase. This limitation is addressed by introducing complex-domain signals, which implicitly include phase comparison with a reference signal, thus enabling both the instantaneous amplitude and phase to be described in a unique and intuitive manner. This approach has found widespread application in modern fields and practical settings, including but not limited to communication systems, radar systems, meteorological observations, and vibration sensing.

Modal decomposition, as a general adaptive filter, has been widely applied for signal analysis and processing in various contexts. For example, Empirical Mode Decomposition (EMD)\cite{1} has been used for rolling bearing fault detection \cite{2}, bridge strain extraction \cite{3}, and offshore vessel identification \cite{4}, and so forth. Variational Mode Decomposition (VMD)\cite{5} has been applied in radar signal processing \cite{6,7} and elbow pipeline monitoring \cite{8}. Symplectic Geometry Modal Decomposition (SGMD)has been utilized for rolling bearing fault detection \cite{9} and photovoltaic power prediction \cite{10}. Recently, we combined the advantages of SGMD and VMD to propose Robust Modal Decomposition (RMD) \cite{11}, which balances numerical optimization methods (e.g., EMD, VMD) and spectral analysis methods (e.g., SSA, SGMD). Compared to these methods, RMD demonstrates superior noise immunity and nonlinear signal processing capabilities. However, existing works have been limited to real-valued data, restricting their practical applicability in signal processing and related fields. Therefore, this paper extends RMD into the complex domain to handle replicated time series data.

In fact, extending modal decomposition methods originally designed for real-valued signals to the complex domain is a natural and general extension, as seen in approaches such as Bivariate Empirical Mode Decomposition (BEMD) \cite{12}, Complex Variational Mode Decomposition (CVMD) \cite{13}, Complex Singular Spectrum Analysis (CSSA) \cite{14}, and Complex Synchronized Generalized Modal Decomposition (CSGMD) \cite{14}. It is important to note that real-world data often contain substantial noise. Compared to all the aforementioned methods, the RMD approach, with similar or even fewer computational demands, achieves superior noise robustness, which is crucial for signal processing applications in real-world scenarios. Excitingly, mathematical derivations and extensive experiments demonstrate that the proposed Complex Robust Modal Decomposition (CRMD) method preserves all the advantages of RMD in terms of noise immunity and other aspects, significantly enhancing its potential for practical applications.

The remainder of this paper is organized as follows: Section 2 reviews the fundamental mathematical principles of RMD and extends them to the complex domain; Section 3 presents the detailed algorithmic framework for CRMD; Section 4 provides numerical experiments and applications to four real-world datasets; finally, conclusions are presented in Section 5.

  \section{MATHEMATICAL PRINCIPLES}

\subsection{Bandwidth Constraints in Complex Domain}
For a complex modal signal \( u_k(t) = A_k(t)e^{j\phi_k(t)} \) (where \( j = \sqrt{-1} \)), the derivative is:
\begin{equation}
u_k'(t) = \left( A_k'(t) + jA_k(t)\omega_k(t) \right) e^{j\phi_k(t)}
\end{equation}
with instantaneous frequency \( \omega_k(t) = \phi_k'(t) \). Under slow amplitude variation (\( A_k'(t) \approx 0 \)):
\begin{equation}
u_k'(t) \approx jA_k(t)\omega_k(t)u_k(t)
\end{equation}
The \( L_2 \)-norm energy constraint becomes:
\begin{equation}
\int |u_k'(t)|^2 dt \approx \int A_k^2(t)\omega_k^2(t) dt
\end{equation}
For discrete signals, this translates to:
\begin{equation}
\|\nabla u_k\|_2^2 = \sum_n |u_k[n+1] - u_k[n]|^2
\label{eq:disc_energy}
\end{equation}

\subsection{Complex-Valued Trajectory Matrix and PCA}
The Hankel trajectory matrix for a complex signal \( x[n] \in \mathbb{C} \) is:
\begin{equation}
\mathbf{X} = \begin{bmatrix}
x[1] & x[2] & \cdots & x[N-K+1] \\
x[2] & x[3] & \cdots & x[N-K+2] \\
\vdots & \vdots & \ddots & \vdots \\
x[K] & x[K+1] & \cdots & x[N]
\end{bmatrix} \in \mathbb{C}^{K \times (N-K+1)}
\end{equation}
where \( K \) is the embedding dimension. The Gram matrix is:
\begin{equation}
\mathbf{G} = \mathbf{X}\mathbf{X}^\dagger \in \mathbb{C}^{K \times K}
\end{equation}
with \( \dagger \) denoting the conjugate transpose. Singular Value Decomposition (SVD) of \( \mathbf{X} \) is:
\begin{equation}
\mathbf{X} = \mathbf{U}\mathbf{\Sigma}\mathbf{V}^\dagger
\end{equation}
where \( \mathbf{U} \in \mathbb{C}^{K \times r} \), \( \mathbf{V} \in \mathbb{C}^{(N-K+1) \times r} \), and \( \mathbf{\Sigma} = \text{diag}(\sigma_1, \dots, \sigma_r) \).

Modal reconstruction uses:
\begin{equation}
\mathbf{X}_i = \sigma_i \mathbf{u}_i \mathbf{v}_i^\dagger
\end{equation}
followed by diagonal averaging to recover 1D components.

\subsection{Regularized Optimization in Complex Domain}
The objective function balancing variance and bandwidth constraint is:
\begin{equation}
\mathcal{J}(\mathbf{w}) = \mathbf{w}^\dagger \mathbf{G}\mathbf{w} - \mu \mathbf{w}^\dagger \mathbf{X}^\dagger \mathbf{R}\mathbf{X}\mathbf{w}
\end{equation}
where \( \mathbf{R} = \mathbf{D}^\dagger \mathbf{D} \) (with \( \mathbf{D} \) as the difference operator). This reduces to a generalized eigenvalue problem:
\begin{equation}
\mathbf{G}\mathbf{v} = \gamma \left( \mathbf{I} + \alpha \mathbf{X}^\dagger \mathbf{R}\mathbf{X} \right) \mathbf{v}
\end{equation}
with \( \gamma \) as the generalized eigenvalue and \( \alpha \) as the regularization parameter.

\subsection{Noise Robustness}
For noisy signals \( x(t) = s(t) + n(t) \) (where \( n(t) \) is complex white noise), the perturbed Gram matrix is:
\begin{equation}
\mathbf{G} = (\mathbf{S} + \mathbf{N})(\mathbf{S} + \mathbf{N})^\dagger = \mathbf{S}\mathbf{S}^\dagger + \mathbf{E}
\end{equation}
where \( \mathbf{E} \) is the noise perturbation. The regularization term suppresses wideband noise by attenuating modes with large \( \mu_i = \mathbf{v}_i^\dagger \mathbf{X}^\dagger \mathbf{R}\mathbf{X}\mathbf{v}_i \), reducing variance while preserving narrowband signals.

\section{Method Overview}

CRMD extends the real-valued RMD to the complex domain, enabling decomposition of complex signals while preserving phase information and enhancing noise robustness. The core idea is to constrain modal bandwidth via regularization in complex phase space, avoiding spurious modes and maintaining physical consistency.

\begin{algorithm}[H]
\caption{Complex Robust Mode Decomposition (CRMD)}
\begin{algorithmic}[1]
\REQUIRE \( \mathbf{x} \in \mathbb{C}^N \), \( r \) (number of modes), \( \theta \) (similarity threshold), \( \alpha \) (regularization factor)
\ENSURE \( \{\mathbf{z}_1,..., \mathbf{z}_r\} \in \mathbb{C}^N \) (modal components), \( \mathbf{z}_{\text{res}} \in \mathbb{C}^N \) (residual)

\STATE Determine \( f_{\text{max}} \) from complex PSD of \( \mathbf{x} \)
\STATE Compute embedding dimension \( K \) using adaptive rule
\STATE Construct complex Hankel matrix \( \mathbf{X} \leftarrow \text{Hankel}(\mathbf{x}, K, \tau=1) \)
\STATE Compute Gram matrix \( \mathbf{G} \leftarrow \mathbf{X}\mathbf{X}^\dagger \)
\STATE Construct difference operator \( \mathbf{D} \) and smoothing matrix \( \mathbf{R} \leftarrow \mathbf{D}^\dagger\mathbf{D} \)
\STATE Set augmented matrix \( \mathbf{M} \leftarrow \mathbf{I} + \alpha\mathbf{R} \)
\STATE Solve generalized eigenvalue problem \( \mathbf{G}\mathbf{v} = \gamma\mathbf{M}\mathbf{v} \)
\STATE Sort eigenvectors \( \{\mathbf{v}_i\} \) by descending eigenvalues \( \gamma \)
\STATE Cluster \( \{\mathbf{v}_i\} \) using complex similarity (\( \text{sim} > \theta \)) to retain \( r \) main eigenvectors
\FOR{$k=1$ to $r$}
    \STATE Reconstruct trajectory matrix \( \mathbf{Z}_k \leftarrow \mathbf{X}\mathbf{v}_k\mathbf{v}_k^\dagger \)
    \STATE Recover modal component \( \mathbf{z}_k \leftarrow \text{diag\_avg}(\mathbf{Z}_k) \)
\ENDFOR
\STATE Compute residual \( \mathbf{z}_{\text{res}} \leftarrow \mathbf{x} - \sum_{k=1}^r \mathbf{z}_k \)

\RETURN \( \{\mathbf{z}_1,..., \mathbf{z}_r\}, \mathbf{z}_{\text{res}} \)
\end{algorithmic}
\end{algorithm}
\clearpage  % 强制输出当前页内容，后续内容（含第二幅图）另起一页
\section{Applications}
\subsection{Synthetic Data}
This section validates the effectiveness of CRMD using artificially synthesized datasets. The signal has a total duration of 10 seconds with a sampling rate of 200 Hz, resulting in 2000 data points. The clean signal is constructed by superimposing three sinusoidal components and one amplitude-modulated (AM) carrier:

\begin{equation}
x_{\text{clean}}(t) = 3\sin(2 \cdot 2\pi t) + 2\cos(7 \cdot 2\pi t) + 5\sin(16 \cdot 2\pi t) + 4\left[1 + 0.5\cos(2\pi t)\right]\cos(32 \cdot 2\pi t)
\end{equation}

\begin{figure}[htbp]
    \centering
    \includegraphics[width=1\linewidth]{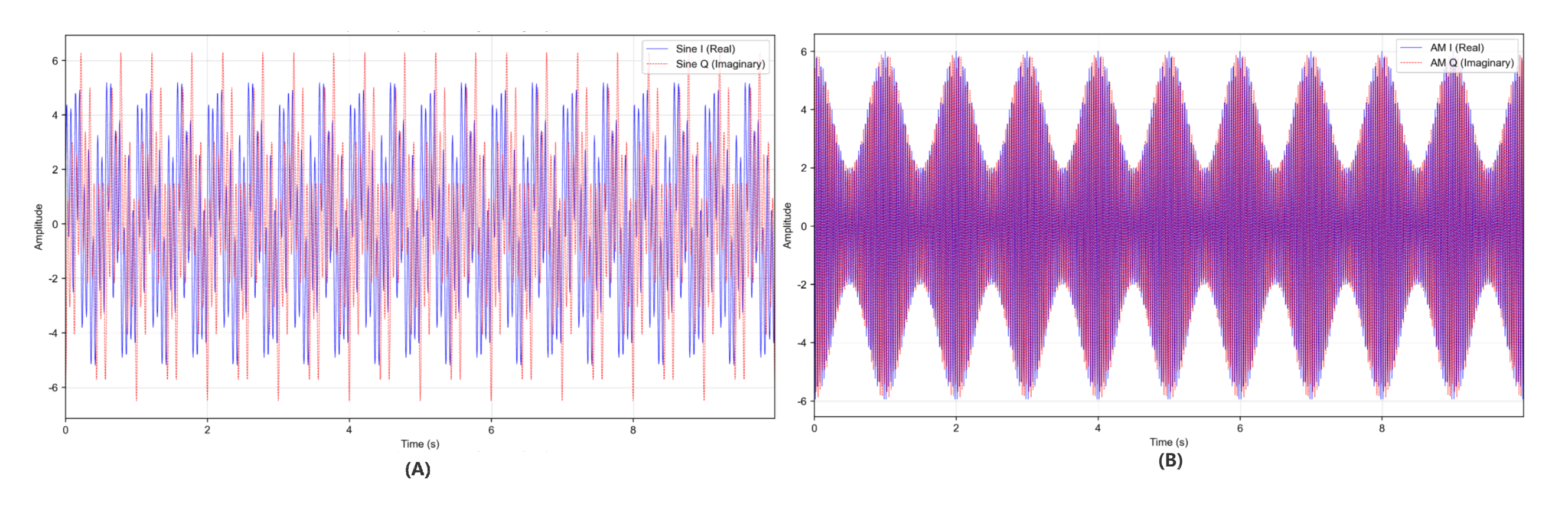}
    \caption{\small presents the individual components of the clean signal. (A) shows the three sinusoidal components with frequencies of 2 Hz, 7 Hz, and 16 Hz; (B) illustrates the AM component with a 32 Hz carrier modulated by a 1 Hz signal. The amplitudes of the components are 3, 2, 5, and 4 units for 2 Hz, 7 Hz, 16 Hz, and AM component, respectively.}
    \label{fig:1}
\end{figure}

\begin{figure}[htbp]
    \centering
    \includegraphics[width=1\linewidth]{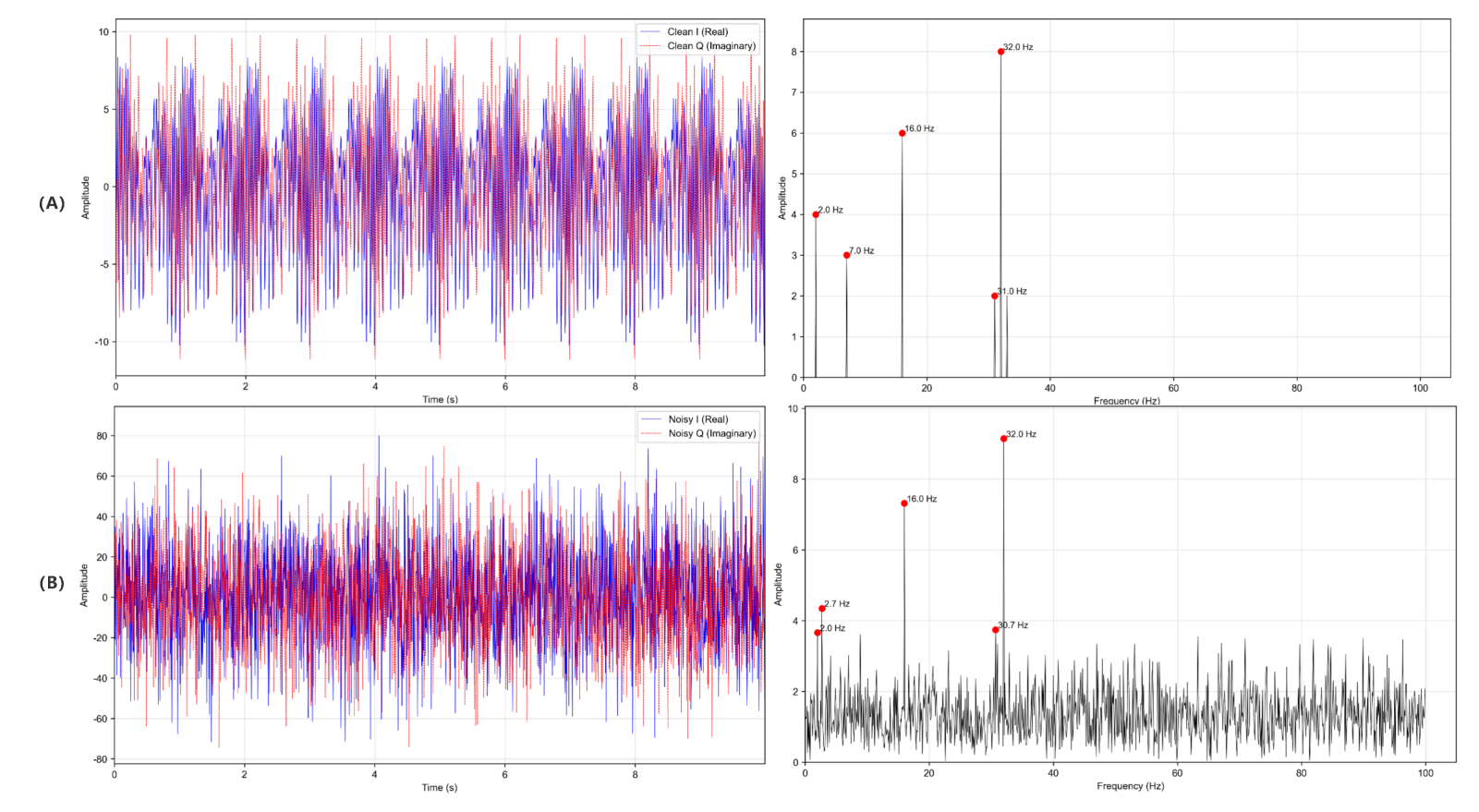}
    \caption{\small presents the time-domain waveforms and spectrograms of the signals used in simulations under the SNR of -15 dB. For better visualization, only the first 2 seconds of the time-domain waveforms are presented. The blue curves correspond to the original clean signal (before noise addition) and its spectrogram, while the red curves represent the signal after noise contamination and its respective spectrogram. As observed from the spectrograms, under the condition of -15 dB SNR, the signal suffers severe degradation—specifically, the 7 Hz and 2 Hz components are almost completely submerged by noise.}
    \label{fig:2}
\end{figure}

\begin{figure}[htbp]
    \centering
    \includegraphics[width=1\linewidth]{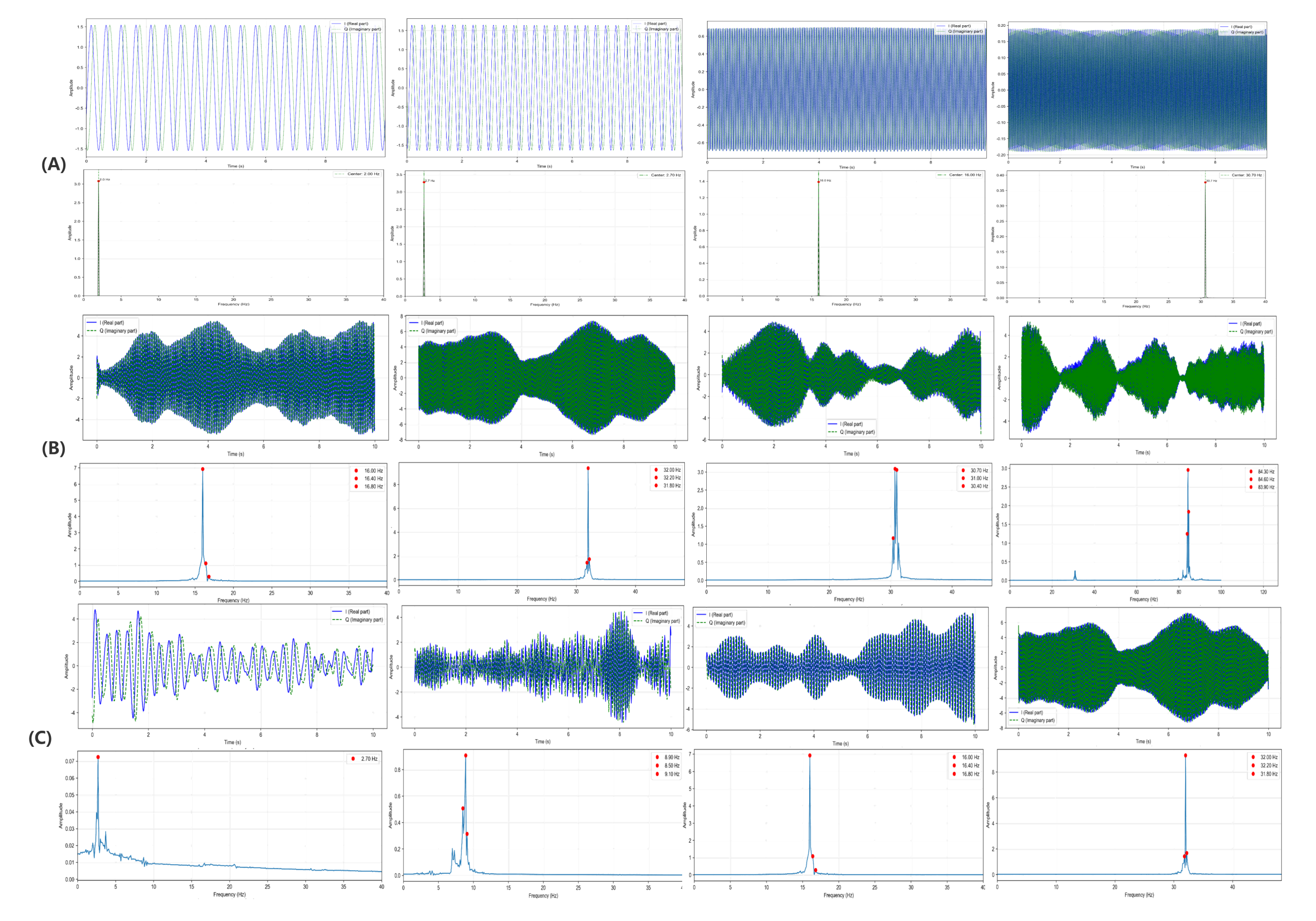}
    \caption{\small presents the modal decomposition results of the three methods under -15 dB SNR. (A) shows the results of CVMD: the spectrum is clean under strong narrowband constraints, but the AM signal is severely distorted, and the 7 Hz sinusoidal component is not separated. (B) illustrates CSGMD results: the amplitude of decomposed modes is distorted, the spectrum is less clean than CVMD, and low-frequency components (2 Hz and 7 Hz) are missing. (C) displays CRMD results: although slight amplitude distortion occurs, no modes are missing—note that the 2 Hz and 7 Hz peaks shift to 2.7 Hz and 8.5 Hz under strong noise, respectively.}
    \label{fig:3}
\end{figure}

The sinusoidal components (Fig. \ref{fig:1}(A)) have frequencies of 2 Hz, 7 Hz, and 16 Hz with amplitudes of 3, 2, and 5 units, respectively. The AM component (see Fig. \ref{fig:1}(B)) has an amplitude of 4 units, a carrier frequency of 32 Hz, and is modulated by a 1 Hz cosine signal with a modulation index of 0.5. Complex Gaussian white noise is then injected to reduce the signal-to-noise ratio (SNR) to -15 dB, simulating a heavily contaminated scenario. This dataset contains both dense spectral lines (2 Hz, 7 Hz, 19 Hz) and broadband AM sidebands, enabling comprehensive evaluation of the algorithm in three aspects: frequency resolution, amplitude preservation, and broadband noise suppression. Fig. \ref{fig:2} presents the time-domain waveforms (including real and imaginary parts) and spectrograms of the clean signal and noisy signal at -15 dB. Comparisons are made with CVMD and CSGMD (when the regularization factor $\alpha = 0$, CRMD is mathematically equivalent to CSGMD). Fig. \ref{fig:3} shows the modal decomposition results of the three methods, including the decomposed time-domain waveforms and spectrograms.

\begin{figure}[htbp]
    \centering
    \includegraphics[width=1\linewidth]{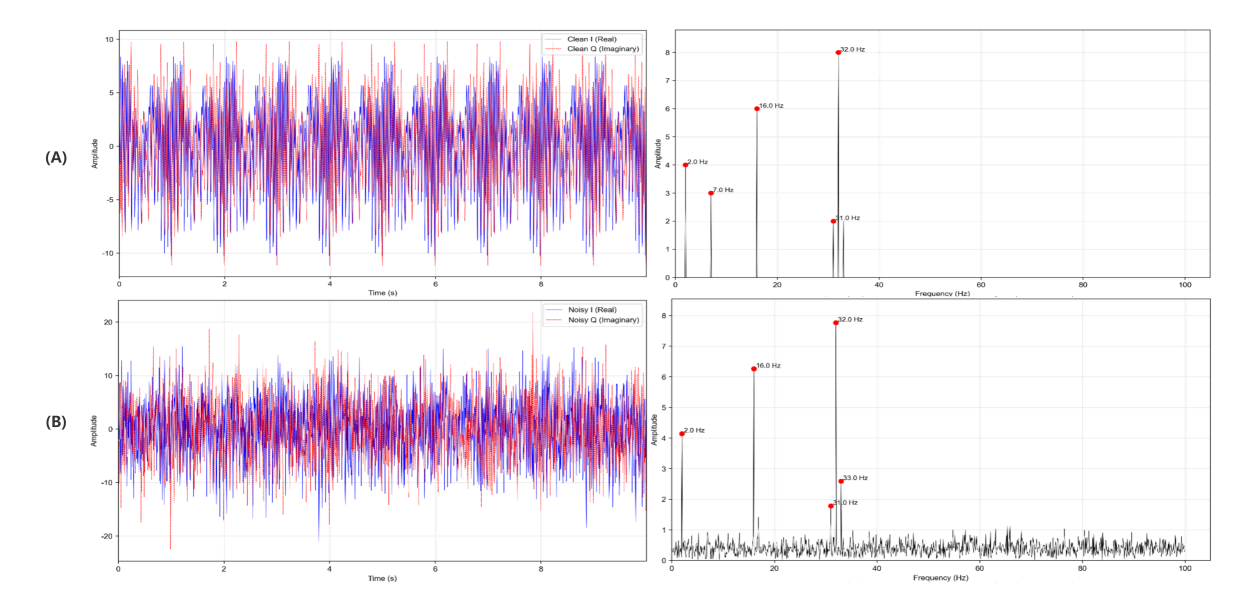}
    \caption{\small presents the time-domain waveforms and spectrograms of the signals under the SNR of -5 dB. Similar to Fig. \ref{fig:2}, the blue curves represent the clean signal and its spectrogram, while the red curves correspond to the noisy signal. With improved SNR, the 16 Hz and AM components become more distinguishable, but the 7 Hz component remains partially obscured by noise.}
    \label{fig:4}
\end{figure}

\begin{figure}[htbp]
    \centering
    \includegraphics[width=1\linewidth]{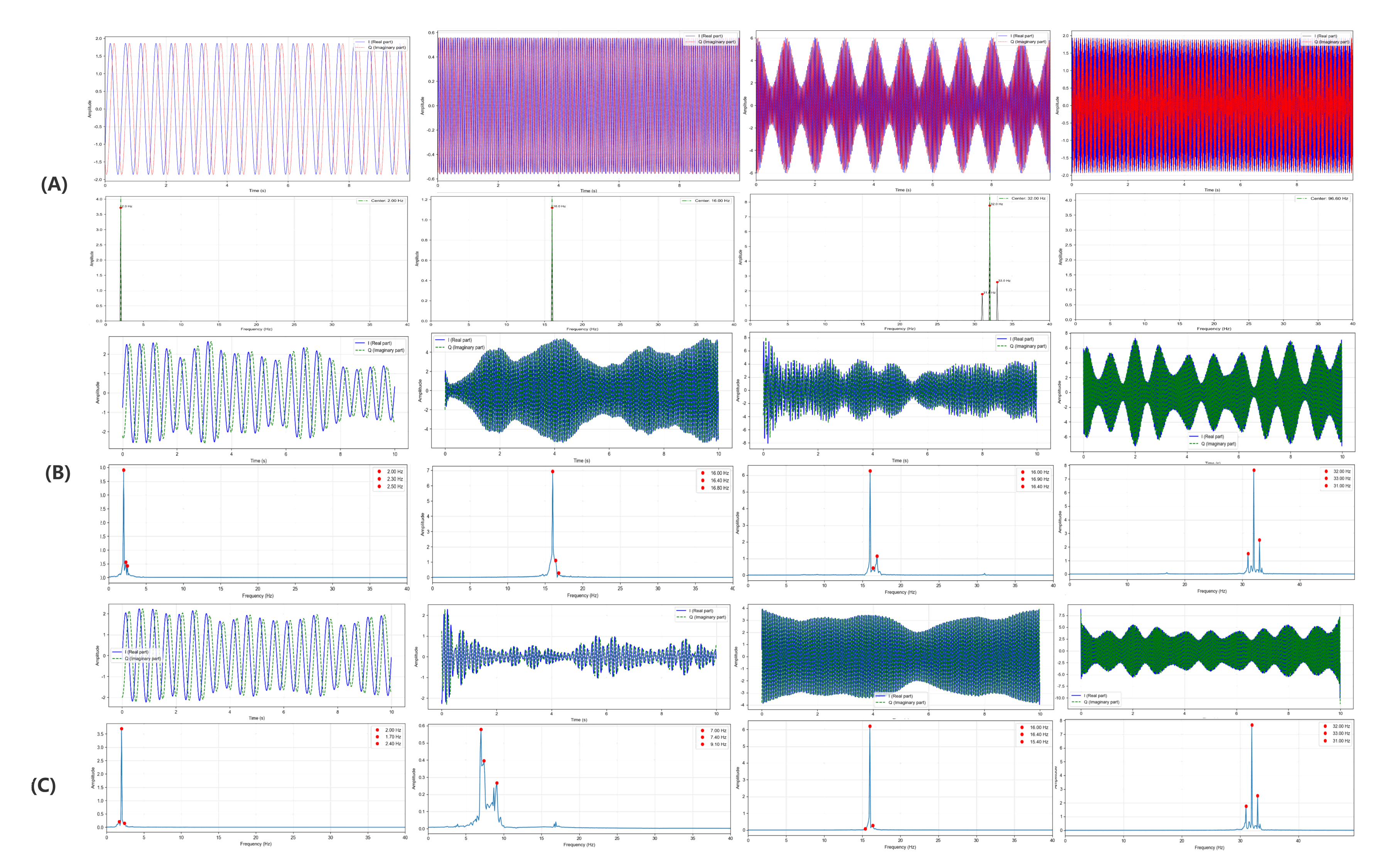}
    \caption{\small presents the modal decomposition results of the three methods under -5 dB SNR. (A) shows CVMD results: the 7 Hz sinusoidal component is still missing, but the decomposed AM component is well-preserved. (B) illustrates CSGMD results: the 7 Hz component is also missing, with noticeable noise residue in the spectrum. (C) displays CRMD results: the only method that successfully identifies the 7 Hz frequency component; despite slight distortion, other components are decomposed more cleanly compared to CSGMD.}
    \label{fig:5}
\end{figure}

Further, we increased the SNR to -5 dB and repeated the experiment. Fig. \ref{fig:4} shows the signal characteristics under this condition, and Fig. \ref{fig:5} presents the corresponding decomposition results.

\clearpage  % 强制输出当前页内容，后续内容（含第二幅图）另起一页

\subsection{Real World Data Experiment}
\subsubsection{Roll Bearing Dataset}
\begin{figure}[H]
    \centering
    \includegraphics[width=0.8\linewidth]{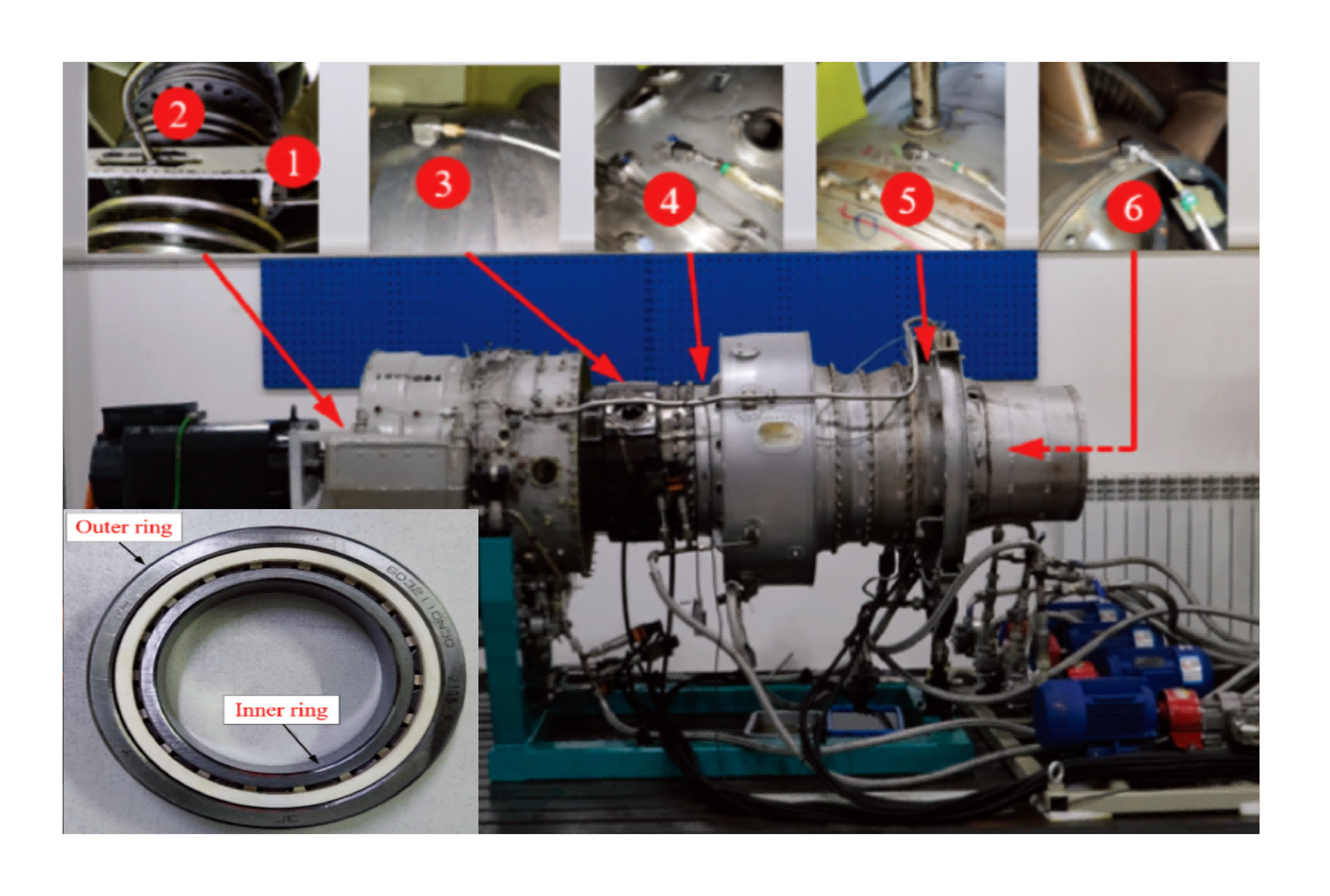} % 适度缩小尺寸
    \caption{\small shows the physical diagram of the bearing fault experimental platform and the bearing diagram from Ref. \cite{16}, with the positions of 6 sensors marked by red arrows.}
    \label{fig:6}
\end{figure}

This section applies the CRMD method to a fault detection dataset of aero-engine rolling bearings, provided by Hou Lei et al. from Harbin Institute of Technology \cite{16}. The dataset includes monitoring data of normal bearings, inner ring fault bearings, and outer ring fault bearings under different rotational speeds of the aero-engine. The physical diagram of the test bench is shown in Fig. \ref{fig:6}, where the positions of each sensor are marked with red arrows. The sensor types include 2 eddy current displacement sensors and 4 vibration sensors; in this dataset, the vertical and horizontal displacement data from the displacement sensors can be treated as complex-valued data, making them suitable for CRMD processing.

\begin{figure}[H]
    \centering
    \includegraphics[width=0.7\linewidth]{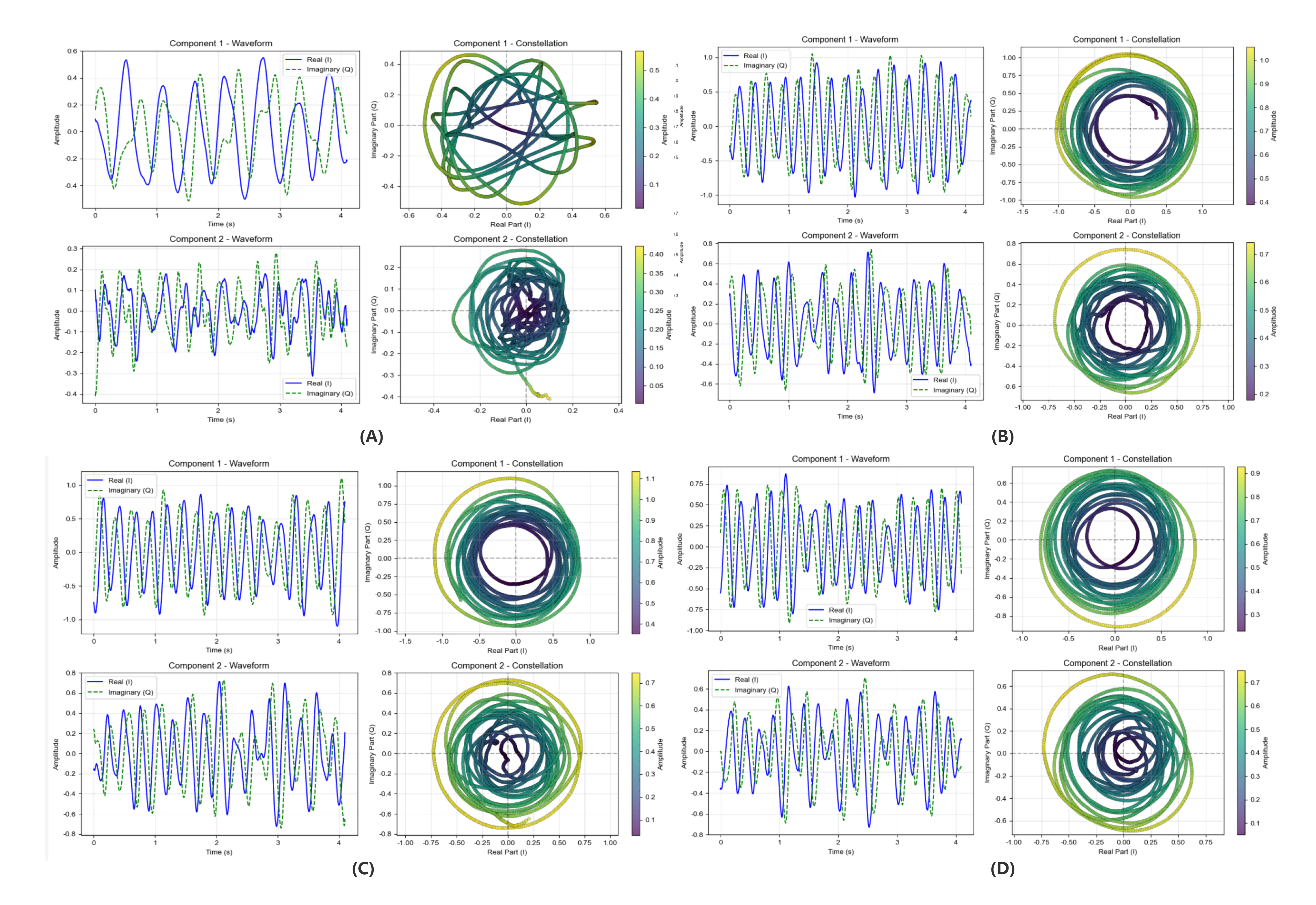}
    \caption{\small presents the experimental results of decomposing displacement sensor data using CRMD. 
    (A) Constellation diagrams of the first two modal components after CRMD decomposition of displacement sensor data from normal bearings; 
    (B), (C), and (D) Constellation diagrams of the first two decomposed modes from displacement sensors of bearings with outer ring cracks, inner ring cracks (0.5 mm), and inner ring cracks (0.1 mm), respectively.}
    \label{fig:7}
\end{figure}

The dataset contains five groups of data: the first two groups are from normal bearings, the third group from an outer ring bearing with a fault crack, and the fourth/fifth groups from inner ring bearings with fault cracks (with crack lengths of 0.5 mm and 1.0 mm, respectively). Each group includes 28 rotational speed combinations. Through CRMD analysis of extensive displacement sensor data, we found that normal bearings hardly decompose into sinusoidal modal components, while faulty bearings typically exhibit two or more sinusoidal modal components. Faulty bearings can thus be detected based on these decomposition results. As shown in previous analyses and Fig. \ref{fig:7}, CRMD outperforms SGMD in anti-interference and noise resistance, which is critical for aero-engine fault monitoring.
\vspace{-6pt}
\vspace{-6pt}
\begin{figure}[H]
    \centering
    \includegraphics[width=0.7\linewidth]{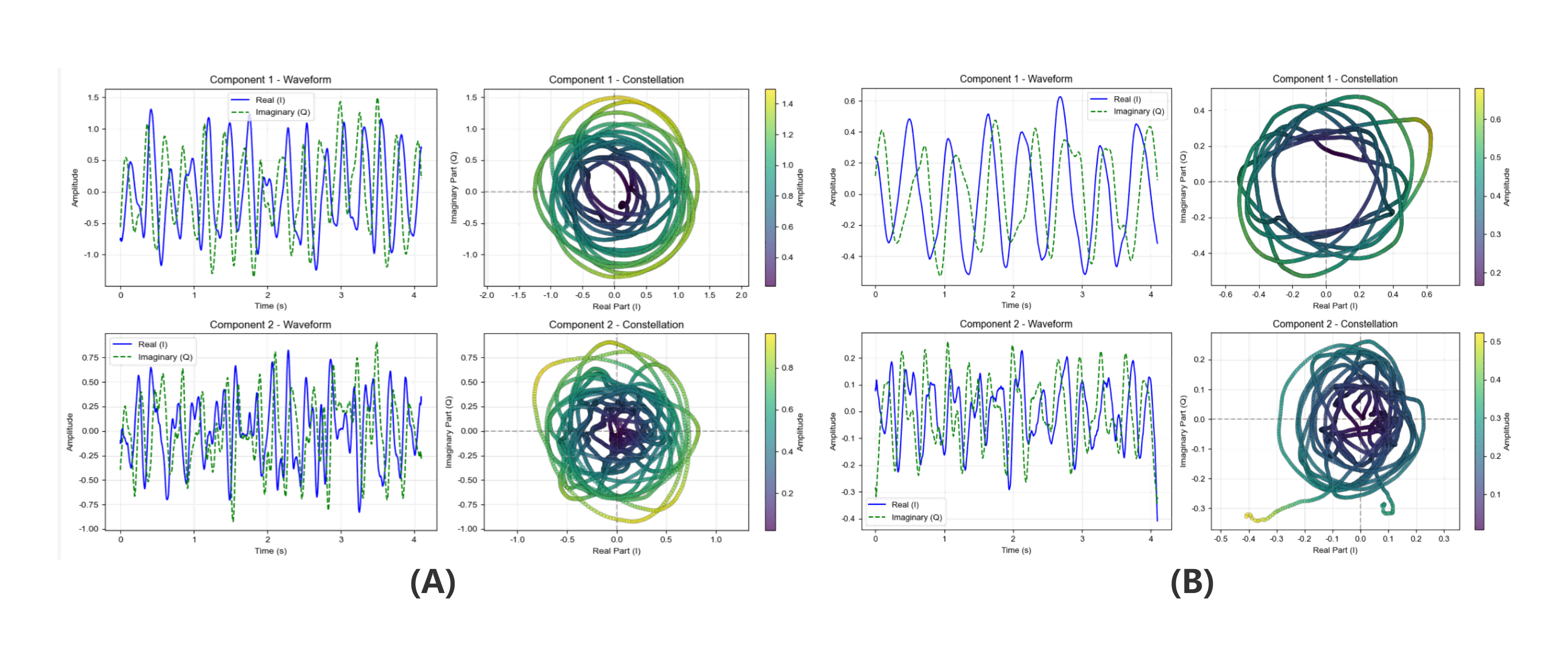}
    \caption{\small displays the decomposition results when $\alpha = 0$. Under high noise or strong interference, fault modal components are submerged in noise, making it easy to confuse abnormal modes with normal modes without careful observation. (A) Modal constellation diagram decomposed from normal bearings; (B) Modal constellation diagram decomposed from faulty bearings.}
    \label{fig:8}
\end{figure}

\subsubsection{Vital Signs Radar Dataset}
\begin{figure}[H]
    \centering
    \includegraphics[width=0.9\linewidth]{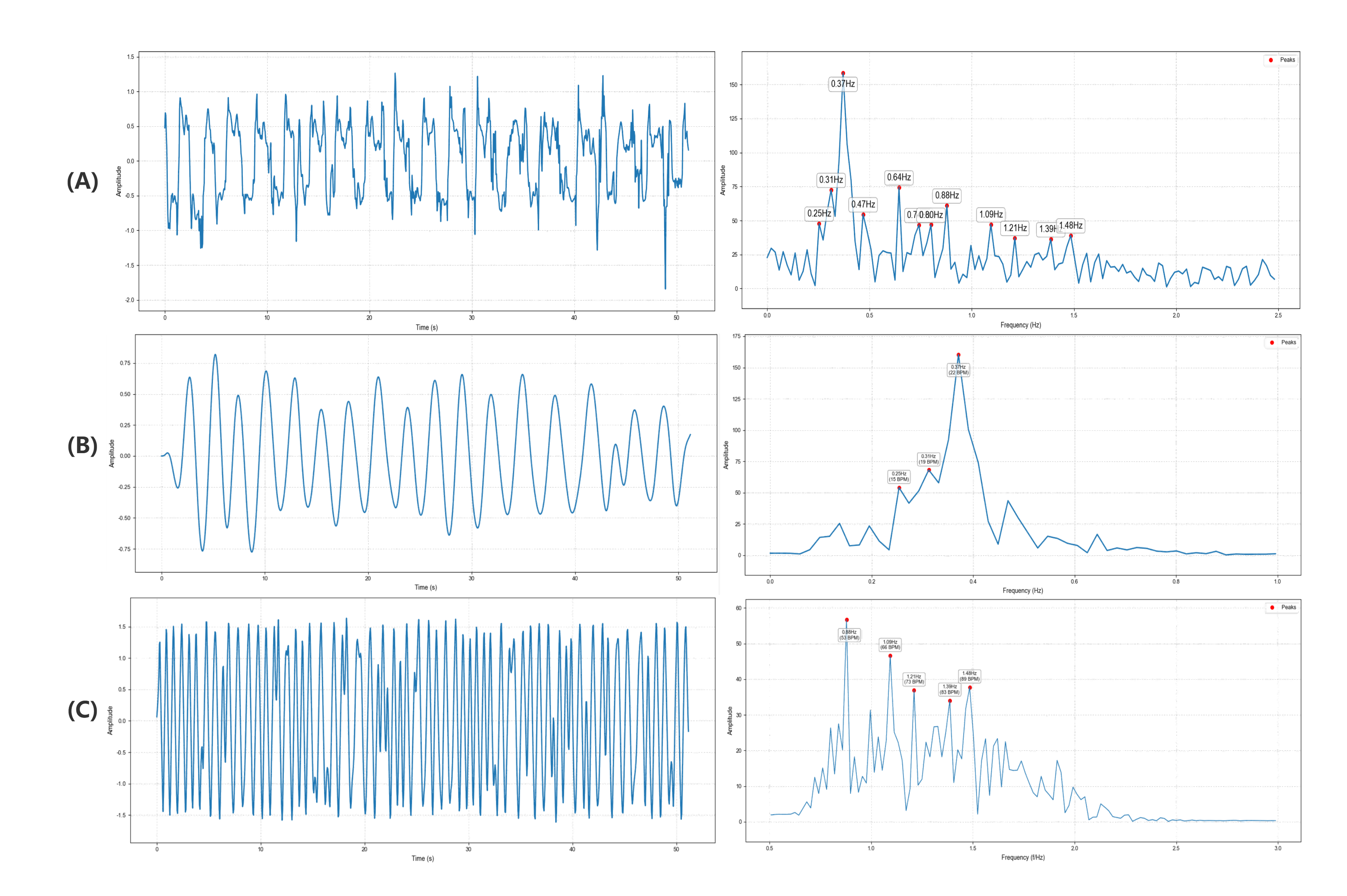}
    \caption{\small shows the respiration and heartbeat waveforms measured by the phase method. 
    (A) Time-domain waveform and its spectrum after phase differencing; (B) Time-domain waveform and its spectrum obtained after filtering with a low-pass filter (0–0.5 Hz); (C) Time-domain waveform and its spectrum obtained after filtering with a band-pass filter (0.5–2 Hz).}
    \label{fig:9}
\end{figure}

\begin{figure}[H]
    \centering
    \includegraphics[width=0.9\linewidth]{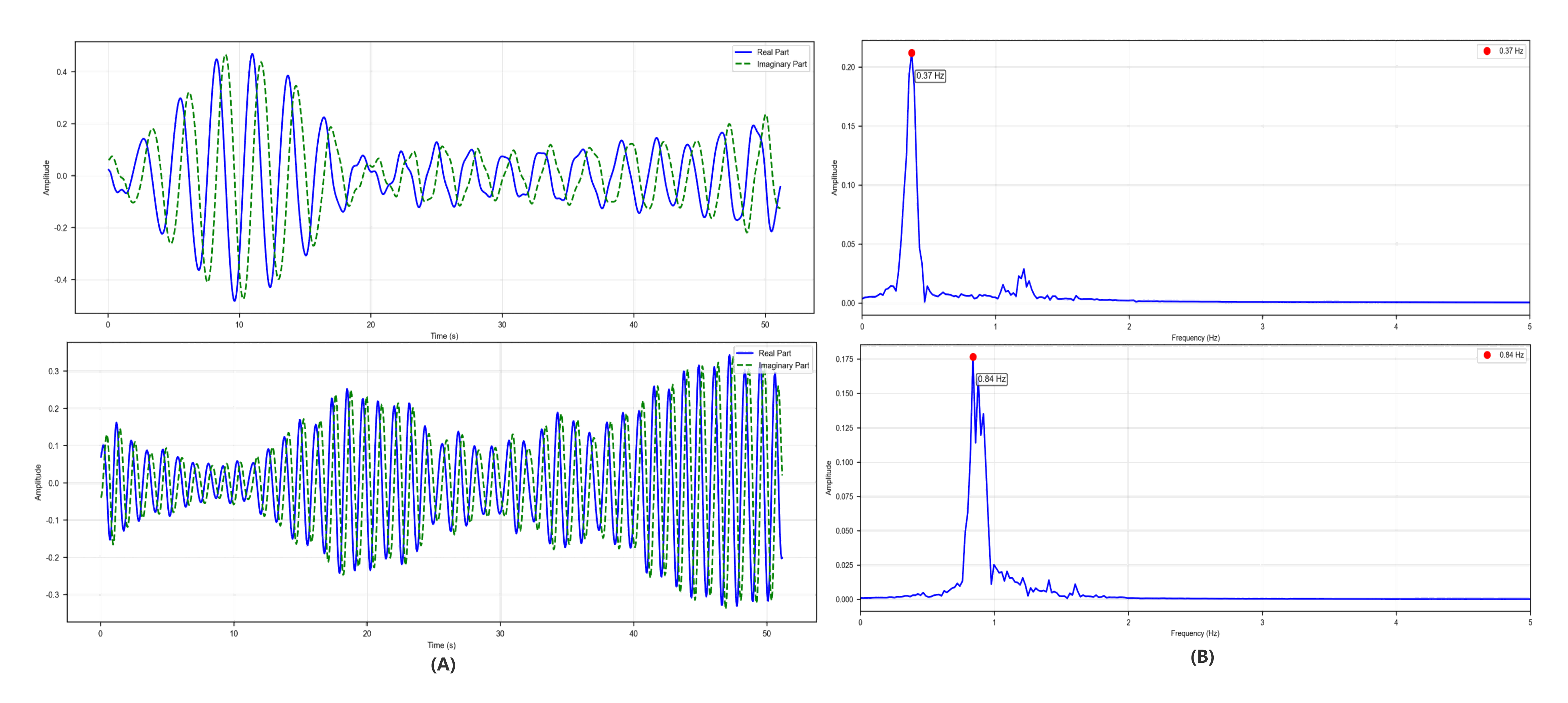}
    \caption{\small presents the respiration and heartbeat waveforms decomposed by CRMD, consistent with the phase method results. 
    Fig. \ref{fig:10} shows the respiration (A) and heartbeat (B) waveforms and their spectra obtained by directly decomposing the analytic signal of phase differencing using CRMD.}
    \label{fig:10}
\end{figure}

This section applies CRMD to human respiration and heartbeat measurement using a millimeter-wave FMCW radar. The dataset was collected by our team using a TI AWR2243 millimeter-wave radar \cite{17}. The original radar echo signals are complex-valued intermediate-frequency IQ signals after de-chirping and orthogonal down-conversion. After performing the first FFT (range-dimensional FFT), complex points within the target range gate at each time point are extracted, yielding slow-time complex signals:

where $d_0$ is the fixed distance and $x(t)$ is the chest micro-motion signal. Consistent with conventional methods \cite{18,19,20,21}, we first extract the phase of the slow-time signal, perform phase inversion and unwrapping, and then directly apply CRMD for decomposition (without further phase differencing or filtering). The resulting respiration and heartbeat waveforms are consistent with those measured by the phase method (Fig. \ref{fig:9}) as shown in Fig. \ref{fig:10}.

\subsubsection{UAV Dataset \label{subsec:uav}}

\begin{figure}[H]
    \centering
    \includegraphics[width=0.9\linewidth]{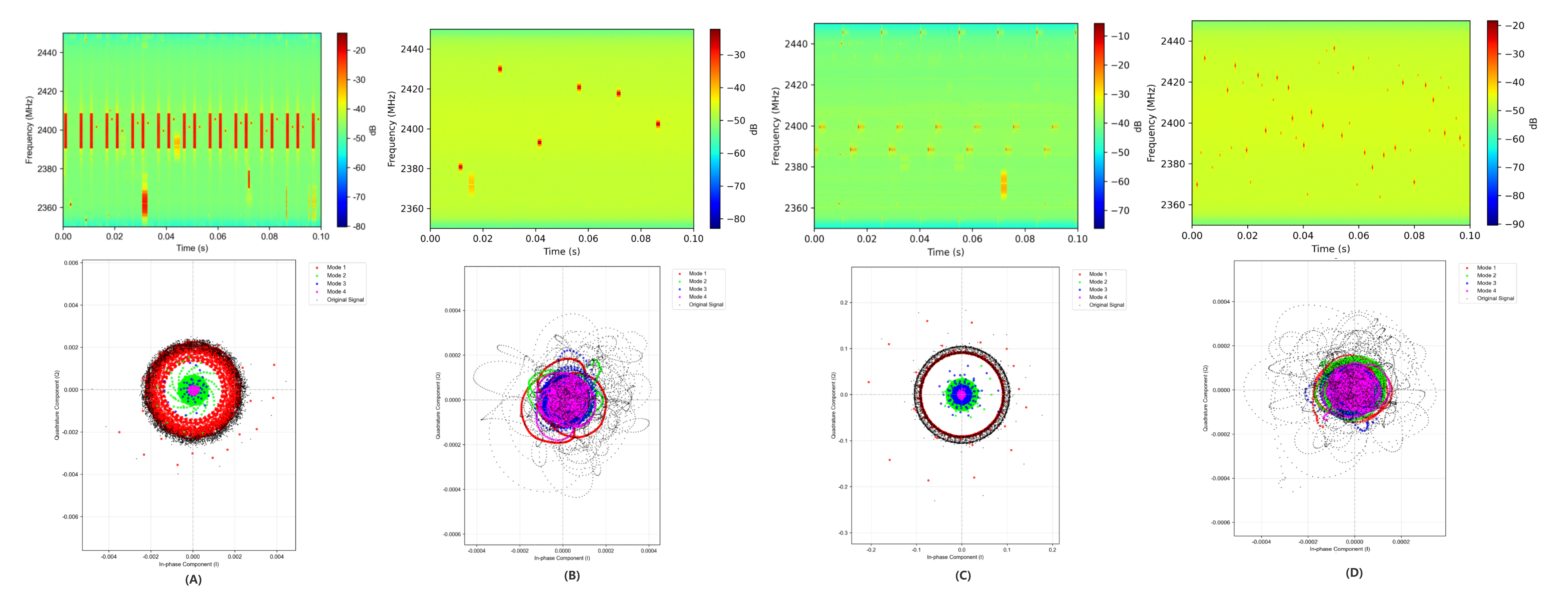}
    \caption{\small demonstrates the time-frequency spectrograms and decomposed constellation diagrams of four UAV RF detection signals. (A) FUTABA T16IZ; (B) DJI MINI3; (C) YunZhuo H12; (D) FLYSKY EL 18. Each subfigure includes the time-frequency spectrogram (top) and the decomposed constellation diagram (bottom).}
    \label{fig:11}
\end{figure}

This section applies the CRMD method to radio frequency (RF) unmanned aerial vehicle (UAV) detection datasets. The UAV dataset used is RFDUAV, collected via the USRP software-defined radio device \cite{22}, which contains RF IQ complex signals of UAVs passing through the scene. RFDUAV includes over 30 types of UAVs. Conventional methods use short-time Fourier transform to generate time-frequency spectrograms (waterfall plots) for identification, while this paper proposes using CRMD to directly decompose constellation diagrams of the first few modes. Different UAVs can be identified based on constellation features, with stronger noise robustness than short-time Fourier transform. Fig. \ref{fig:11} demonstrates the time-frequency spectrograms and decomposed constellation diagrams of four UAV RF signals, corresponding to models FUTABA T16IZ (A), DJI MINI3 (B), YunZhuo H12 (C), and FLYSKY EL 18 (D).
\subsubsection{WiFi CSI for Respiration Measurement \label{subsec:wifi_csi}}

\begin{figure}[H]
    \centering
    \includegraphics[width=0.9\linewidth]{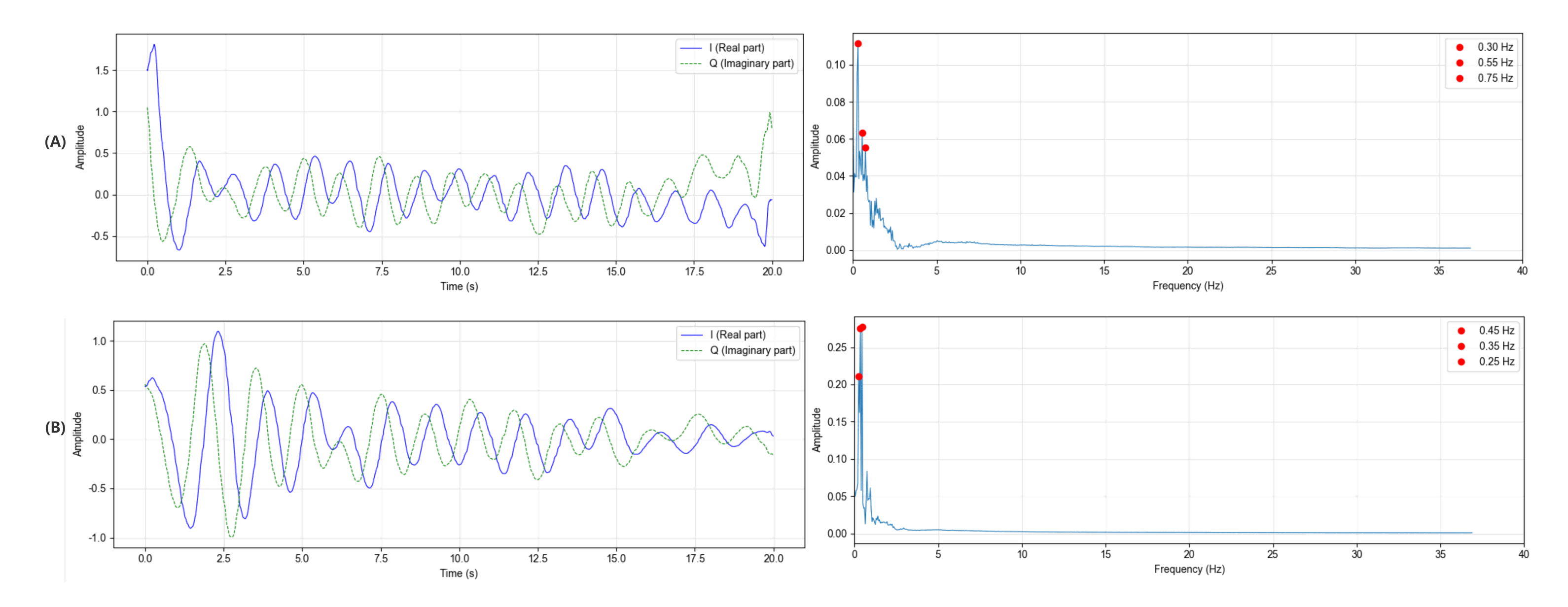}
    \caption{\small shows the respiration signal separated by CRMD and the reference signal from the chest belt sensor. (A) Reference respiration waveform (chest belt sensor) and its frequency (0.3 Hz); (B) Respiration waveform separated by CRMD and its frequency (0.3–0.35 Hz).}
    \label{fig:12}
\end{figure}

\begin{figure}[H]
    \centering
    \includegraphics[width=0.9\linewidth]{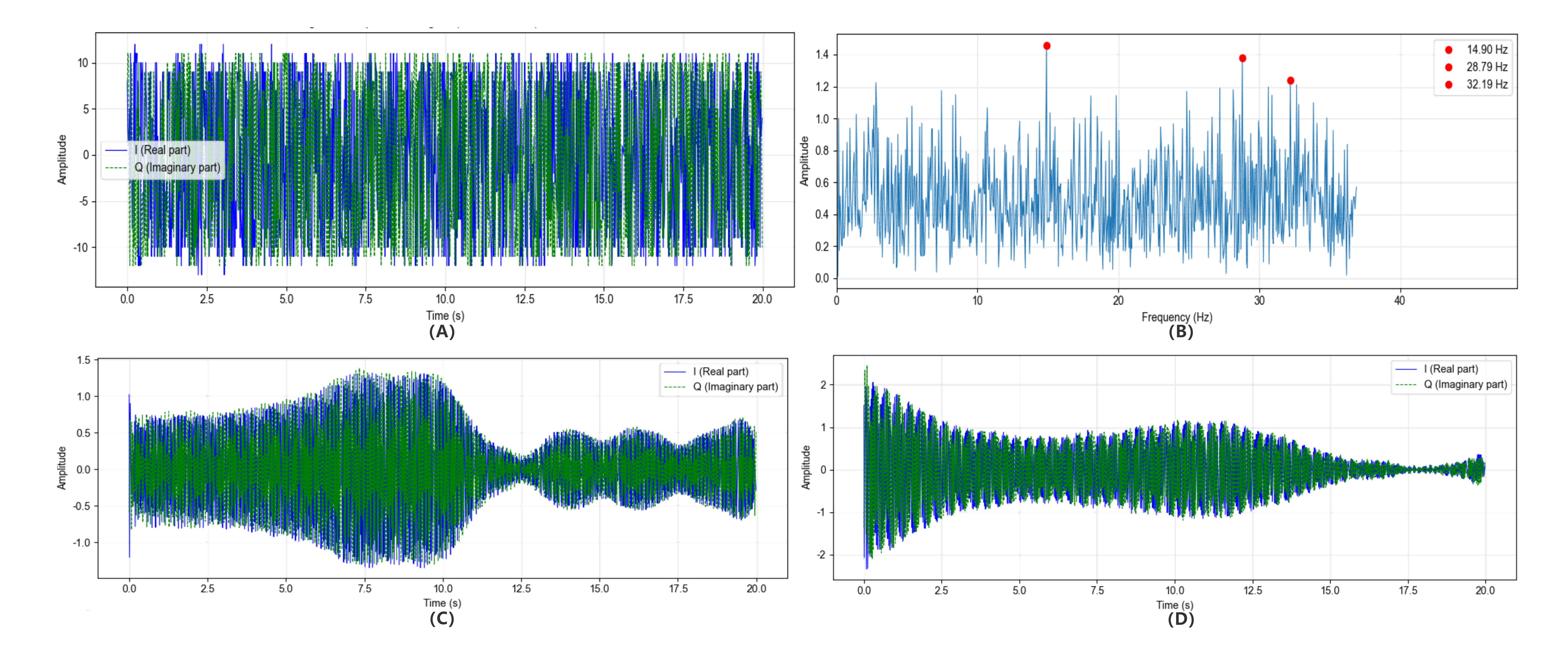}
    \caption{\small presents the decomposition results without constraining the modal bandwidth. (A) and (B) show the original WiFi CSI time-domain waveform and spectrum, respectively, with very low SNR; (C) and (D) are the results when the bandwidth is not constrained ($\alpha=0$), where correct respiratory modal components cannot be separated at all. No valid respiration signals can be separated due to severe noise interference, with only disordered noise components observed.}
    \label{fig:13}
\end{figure}

This section applies the CRMD method to WiFi CSI respiration detection. The dataset used is WiFi CSI signals collected by our team using ESP32 \cite{23}. The signal is heavily mixed with noise, so only CRMD can effectively separate the respiration signal. The respiration frequency measured by a reference chest belt sensor is approximately 18 breaths per minute (equivalent to 0.3 Hz), while the frequency of the respiration waveform separated by CRMD is around 0.3–0.35 Hz, which is within an acceptable error range (Fig. \ref{fig:12}). Without constraining the modal bandwidth, no useful signals can be separated (Fig. \ref{fig:13}).

\section{Conclusion}
This paper introduces the CRMD method, which extends the bandwidth-constrained Robust Modal Decomposition (RMD) from the real domain to the complex domain. The feasibility of this extension is elaborated from a mathematical perspective, and extensive experiments validate the advantages of CRMD in noise resistance and other aspects.
In future research, we will more comprehensively and meticulously apply RMD and CRMD to various datasets to fully explore their applicability and limitations in low-signal-to-noise ratio (SNR) and non-stationary signal analysis in the complex domain. For instance, issues such as adaptive parameter adjustment and endpoint effects in the current RMD method require further analysis. Additionally, we have initiated work on computational optimization and embedded deployment of the algorithm. It is hoped that this work will assist researchers in more fields to better analyze and process complex-domain signals.

\bibliographystyle{unsrt} % 数字编号的参考文献样式
\bibliography{sample} % 关联你的.bib文件
\end{document}